\documentclass[twocolumn,showpacs,preprintnumbers,amsmath,amssymb]{revtex4}
\usepackage{graphicx}
\usepackage{epsfig}
\usepackage{dcolumn}
\usepackage{bm}

\def\pA{proton-nucleus\ }
\def\aA{$\alpha$-nucleus\ }
\def\AA{nucleus-nucleus\ }
\def\phe6{$^6$He+$p$\ }
\def\he6pn{$p(^6$He,$^6$Li$^*)n$\ }

\def\pphe6{$p(^6$He,$^6$He)$p$\ }
\def\p2he6{$p(^6$He,$^6$He$^*)p$'\ }
\def\nli6{$^6$Li$^*+n$\ }
\begin{document}
\title{Isospin dependence of \phe6 optical potential and the symmetry energy}
\author{Dao T. Khoa}\email{khoa@vaec.gov.vn}
\author{Hoang Sy Than}
\affiliation{Institute for Nuclear Science {\rm \&} Technique, VAEC, P.O. Box
5T-160, Nghia Do, Hanoi, Vietnam.}

\date{Febr. 18th, 2005, accepted for publication in Phys. Rev. C}
\begin{abstract}
A consistent folding analysis of the elastic \pphe6 scattering and charge
exchange \he6pn\ reaction data measured at $E_{\rm lab}=41.6A$ MeV has been
performed within the coupled channels formalism. We have used the isovector
coupling to link the isospin dependence of \phe6 optical potential to the cross
section of \he6pn reaction exciting the 0$^+$ isobaric analog state (IAS) at
3.563 MeV in $^6$Li. Based on these results and the Hartree-Fock calculation of
asymmetric nuclear matter using the same isospin-dependent effective
nucleon-nucleon interaction, we were able to confirm that the most realistic
value of the symmetry energy $E_{\rm sym}$ is around 31 MeV. Our analysis has
also shown that the measured charge exchange \he6pn\ data are quite sensitive to
the halo tail of the $^6$He density used in the folding calculation and the IAS
of $^6$Li is likely to have a halo structure similar to that established for the
ground state of $^6$He.
\end{abstract} \pacs{24.50.+g, 25.60.Bx, 25.60.Lg,} \maketitle

The knowledge about the symmetry part of the nuclear equation of state (EOS) is
vital for the understanding of the dynamics of supernovae explosion and the
formation of neutron stars \cite{Bet90,Swe94}. The symmetry part of the nuclear
EOS is actually determined by the nuclear matter (NM) symmetry energy $S(\rho)$
defined in terms of a Taylor series expansion of the NM binding energy
$B(\rho,\delta)$ as
\begin{equation}
B(\rho,\delta)=B(\rho,0)+S(\rho)\delta^2+O(\delta^4)+...
 \label{e1}
\end{equation}
where $\delta=(\rho_n-\rho_p)/\rho$ is the neutron-proton asymmetry parameter.
The contribution of $O(\delta^4)$ and higher-order terms in Eq.~(\ref{e1}),
i.e., the deviation from the parabolic law was proven to be negligible
\cite{Kho96,Zuo99}. The NM symmetry energy determined at the NM saturation
density, $E_{\rm sym}=S(\rho_0)$ with $\rho_0\approx 0.17$ fm$^{-3}$, is widely
known in the literature as the \emph{symmetry energy} or symmetry coefficient.
Although numerous nuclear many-body calculations have predicted $E_{\rm sym}$ to
be around 30 MeV (see, e.g., Refs.~\cite{Kho96,Zuo99,Bra85,Pea00,Die03}), a
direct experimental determination of $E_{\rm sym}$ still remains a challenging
task. One needs, therefore, to relate $E_{\rm sym}$ to some experimentally
inferrable quantity like the neutron skin in neutron-rich nuclei
\cite{Bro00,Hor01,Fur02,Die03} or the fragmentation data of heavy-ion (HI)
collisions involving $N\neq Z$ nuclei \cite{Bao02,She04,Che04}. An accurate
estimate of the $E_{\rm sym}$ value is also very important for the nuclear
astrophysics. For example, a small variation of $E_{\rm sym}$, used as input for
the hydrodynamic simulation of supernovae, affects significantly the electron
capture rate in the ``prompt" phase of type II supernovae \cite{Swe94}. Another
example is a calculation of NM and masses of finite nuclei using Skyrme forces
\cite{Pea00} which shows that the neutron-rich NM does not collapse only if the
corresponding $E_{\rm sym}$ value is within the range $28-31$ MeV. $E_{\rm sym}$
is also an important input for the study of the density dependence $S(\rho)$
based on transport-model simulation of the HI collisions (see Ref.~\cite{Bao02}
and references therein), and the most recent transport-model results favor
$E_{\rm sym}\approx 31-32$ MeV \cite{Che04}.

Within the frame of any microscopic model for asymmetric NM, the symmetry energy
depends strongly on the isospin dependence of the nucleon-nucleon ($NN$)
interaction used therein \cite{Kho96,Zuo99}. Therefore, the $E_{\rm sym}$ value
can be indirectly tested in a charge exchange (isospin-flip) reaction which has
been known for decades as a good probe of the isospin dependence of the
effective $NN$ interaction \cite{Doe75}. Although the isospin dependence of the
optical potential (OP), known by now as Lane potential \cite{La62}, has been
studied since a long time, there has been a considerable interest recently in
studying the isospin dependence of the OP in the quasi-elastic scattering
reactions measured with unstable neutron-rich beams. Based on the isospin
symmetry, the \AA OP can be written in terms of an isovector coupling
\cite{La62} as
\begin{equation}
 U(R)=U_0(R)+4U_1(R)\frac{{\bm t}.{\bm T}}{aA}, \label{e2}
\end{equation}
where ${\bm t}$ is the isospin of the projectile $a$ and ${\bm T}$ is that of
the target $A$. For a proton-induced scattering reaction, the second term of
Eq.~(\ref{e2}) contributes to both the elastic ($p,p$) scattering and ($p,n$)
charge exchange reaction \cite{Sat83}. While the relative contribution by the
Lane potential $U_1$ to the elastic ($p,p$) cross section is small and amounts
only a few percent for a neutron-rich target \cite{Kho02,Kho03}, it determines
entirely the (Fermi-type) $\Delta J^\pi=0^+$ transition strength of the ($p,n$)
reaction leading to an isobaric analog state (IAS). Therefore, the ($p,n$)
reaction has been so far the main tool in studying the isospin dependence of the
\pA OP. Since this isospin dependence should be better tested in the charge
exchange reactions induced by the neutron-rich beams, we consider in the present
work the \he6pn reaction measured by Cortina-Gil {\sl et al.} \cite{Gil98} with
the secondary $^6$He beam at $E_{\rm lab}=41.6A$ MeV. Given a large
neutron-proton asymmetry ($\delta=1/3$) of the unstable $^6$He nucleus, the
measured \he6pn cross section for the transition connecting the ground state
(g.s.) of $^6$He ($T=T_z=1$) and its isobaric analog partner ($T=1, T_z=0,
J^\pi=0^+$ excited state of $^6$Li at 3.563 MeV) is indeed a good probe of the
isovector coupling in the \phe6 system. In the two-channel approximation, the
elastic \pphe6 scattering and charge exchange \he6pn cross sections can be
obtained from the solutions of the following coupled channels (CC) equations
\cite{Sat83}
\begin{eqnarray}
\left[K_p+U_p(R)-E_p\right]
 \chi_p({\bm R})=-\frac{\sqrt{2}}{3}U_1(R)\chi_n({\bm R}), \label{e3}\\
\left[K_n+U_n(R)-E_n\right]
 \chi_n({\bm R})=-\frac{\sqrt{2}}{3}U_1(R)\chi_p({\bm R}).
 \label{e4}
\end{eqnarray}
Here $K_{p(n)}$ and $E_{p(n)}$ are the kinetic-energy operators and
center-of-mass energies of the \phe6 and \nli6 channels, with the energy shift
due to the Coulomb energy and $Q$-value of the \he6pn reaction properly taken
into account. $U_p(R)$ is the OP in the \phe6 channel and $U_n(R)$ is that in
the \nli6 channel. In addition to the charge exchange \he6pn \cite{Gil98} and
elastic \pphe6 scattering \cite{Gil97} data measured at $41.6A$ MeV, a total
reaction cross section $\sigma_{\rm R}=409\pm 22$ mb was also measured
\cite{Vis01} for the \phe6 system at a slightly lower energy of $36A$ MeV. Thus,
these data sets are the important constraints for the \phe6 OP at the considered
energy.

To link the Lane potential $U_1$ to the isospin dependence of the $NN$
interaction, we have used the folding model \cite{Kho02,Kho03} to calculate
$U_0$ and $U_1$ using the explicit proton and neutron g.s. densities of $^6$He
and the CDM3Y6 density- and isospin dependent $NN$ interaction \cite{Kho97}.
This interaction is based on the M3Y interaction $v_{0(1)}(s)$ deduced from the
G-matrix calculation \cite{Ana83} using the Paris $NN$ potential, with the
energy- and density dependences included explicitly as
\begin{equation}
 v_{0(1)}(E,\rho,s)=(1-0.0026E)F_{0(1)}(\rho)v_{0(1)}(s),
\label{g1}
\end{equation}
where $E$ is the bombarding energy (per projectile nucleon),
\begin{equation}
 F_{0(1)}(\rho)=C_{0(1)}[1+\alpha\exp(-\beta\rho)-\gamma\rho],
\label{g2}
\end{equation}
and the explicit expression of the finite-range $v_{0(1)}(s)$ interactions is
given in Ref.~\cite{Kho96}. Parameters of the \emph{isoscalar} part $F_0(\rho)$
of the density dependence (\ref{g2}) were chosen \cite{Kho97} to reproduce
saturation properties of the symmetric NM in the Hartree-Fock (HF) approximation
and tested in the folding analyses \cite{Kho97,Kho95} of the elastic
\emph{refractive} \AA and \aA scattering to infer realistic estimate for the
nuclear incompressibility [$K(\rho_0)\approx 250$ MeV]. In a similar manner, we
try now to probe the \emph{isovector} part $F_1(\rho)$ in the CC analysis of the
charge exchange \he6pn reaction and, using the HF method of Ref.~\cite{Kho96},
to accurately estimate the symmetry energy $E_{\rm sym}$. Since the CDM3Y6
interaction is real, the folding model is used to calculate the real parts of
the OP, $V_{0(1)}$=Re $U_{0(1)}$, which is further scaled by a complex factor to
obtained $U_{0(1)}$. By taking isospin coupling explicitly into account, one
obtains from Eq.~({\ref{e2})
\begin{eqnarray}
 U_p(R) & = & \left[V_0(R)-\frac{V_1(R)}{3}\right]
 (N_{\rm R}+iN_{\rm I}), \label{g3}\\
 U_n(R) & = & V_0(R)(N_{\rm R}+iN_{\rm I}). \label{g4}
\end{eqnarray}
In the CC calculation, $U_{p(n)}$ are each added by a spin-orbital potential
whose parameters were fixed as taken from the systematics by Becchetti and
Greenless \cite{BG69}, and $U_p$ is added further by a Coulomb potential between
a point charge and a uniform charge distribution of radius $R_{C}=1.35\ A^{1/3}$
fm. The form factor (FF) of the \he6pn reaction, to be used in the right-hand
sides of Eqs.~(\ref{e3}) and (\ref{e4}), is given by Eq.~({\ref{e2}) as
\begin{equation}
 U_{pn}(R)=\frac{2\sqrt{N-Z}}{A}U_1(R)=\frac{\sqrt{2}}{3}V_1(R)
 [1+i(N_{\rm I}/N_{\rm R})].
\label{g5}
\end{equation}
Thus, the scaling factors $N_{\rm R(I)}$ of the real and imaginary parts of the
OP are the main parameters to be determined from the CC description of the
charge exchange \he6pn and elastic scattering \pphe6 data which should also be
\emph{constrained} by a total reaction cross section $\sigma_{\rm R}\approx 400$
mb (an empirical value expected at the considered energy \cite{Vis01}). To have
as less as possible free parameters, we have used a simple assumption (\ref{g5})
to scale the real folded ($p,n$) FF by the same \emph{relative} complex strength
as that used in the elastic channel \cite{Sat79}. Our only nuclear structure
input is the $^6$He$_{\rm g.s.}$ density and we have considered in this work two
different choices: the microscopic density given by the cluster-orbital shell
model approximation (COSMA) \cite{Zhu93,Kor97} and that obtained recently
\cite{Kho04} based on the independent particle model (IPM). The CC calculation
was done with the nonrelativistic code FRESCO \cite{Tho88} using the inputs for
mass numbers and incident energies given by the relativistically corrected
kinematics \cite{Far84}. For a checking purpose, the CC results plotted in
Fig.~\ref{f1} were also compared with those given by the code ECIS97
\cite{Ray97} (which takes exactly into account the relativistic kinematics) and
the two sets of calculated \he6pn cross sections turned out to be nearly
identical.

We found that $N_{\rm R}\approx 0.85$ and $N_{\rm I}\approx 0.55$ which were
mainly determined by the fit to the elastic scattering data and by a constraint
that the calculated $\sigma_{\rm R}$ is around 400 mb. It is noteworthy that a
continuum-discretized coupled channels (CDCC) calculation of the elastic \phe6
scattering at about the same energy by Mackintosh and Rusek \cite{Mac03} has
shown that the real dynamic polarization potential due to $^6$He breakup is
repulsive in the center and at the surface, so that a renormalization factor
$N_{\rm R}<1$ of the real folded OP is well expected. The relative strength
$N_{\rm I}/N_{\rm R}\approx 0.65$ (used further to calculate FF for the \he6pn
reaction) also agrees reasonably with the CDCC results \cite{Mac03} which give
the ratio of volume integrals of the imaginary and real parts of the \phe6 OP
around 0.6.

\begin{figure}[htb]
 \vspace*{-1cm}
 \mbox{\epsfig{file=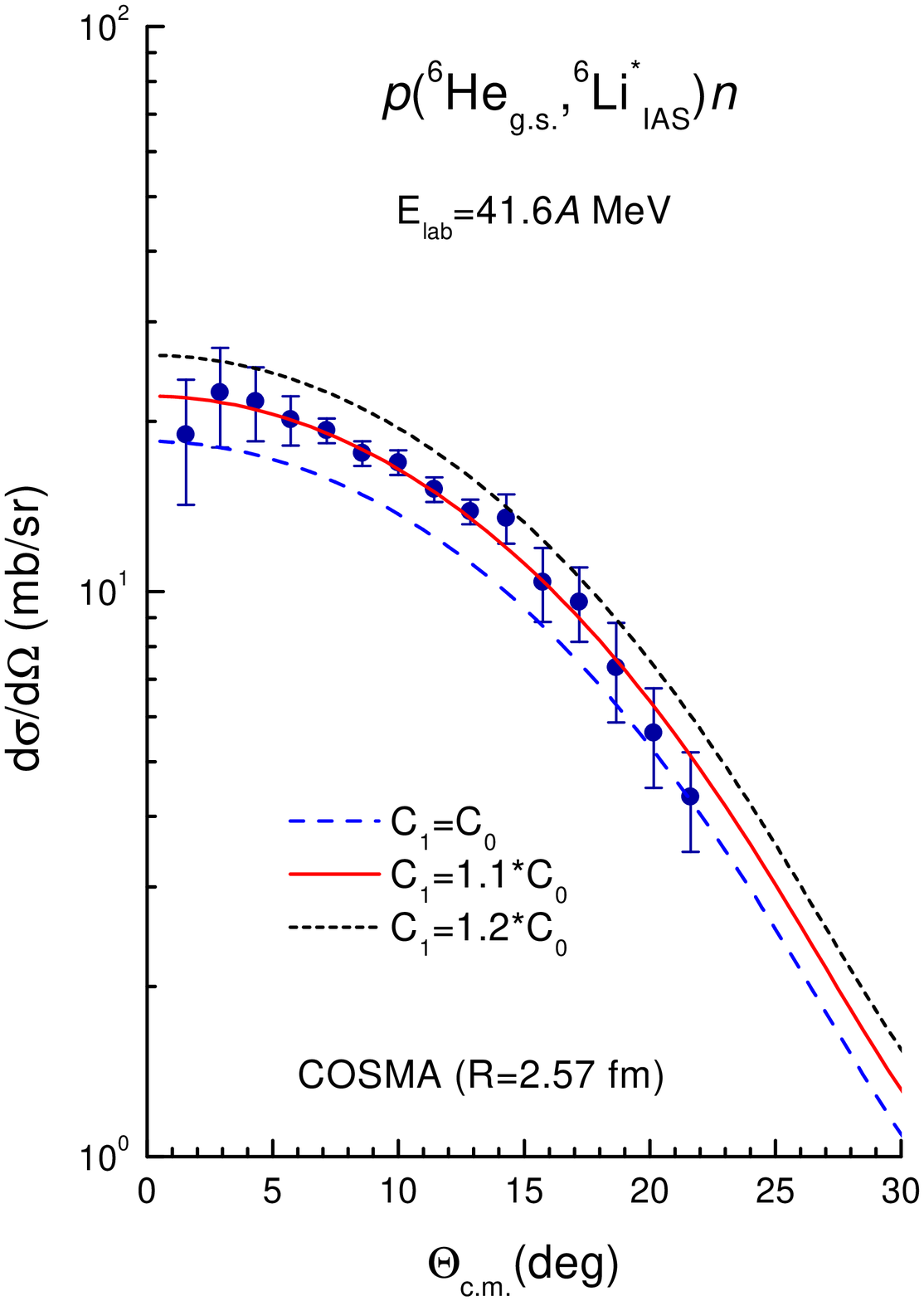,height=12cm}}\vspace*{-1cm}
\caption{CC results for the charge exchange \he6pn cross section at $E_{\rm
lab}=41.6A$ MeV in comparison with the data measured by Cortina-Gil {\sl et
al.} \cite{Gil98}.} \label{f1}
\end{figure}

\begin{figure}[htb]
 \vspace*{-1cm}
 \mbox{\epsfig{file=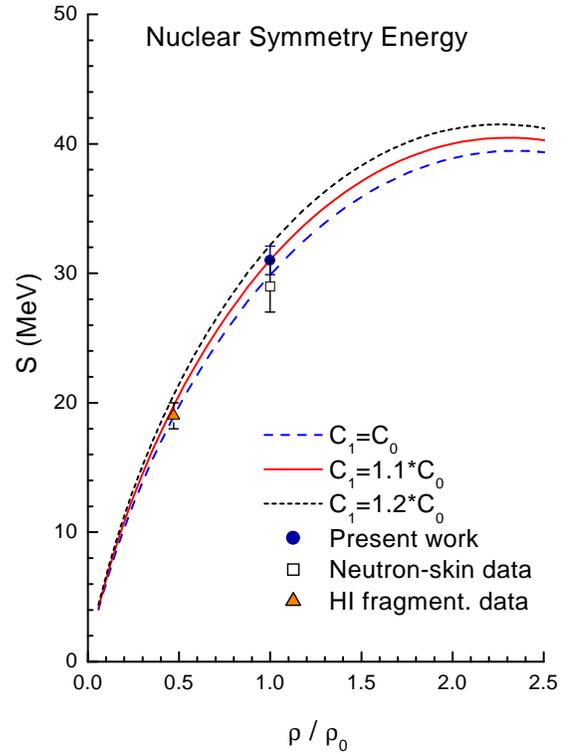,height=12cm}}\vspace*{-1cm}
\caption{Density dependence of the NM symmetry energy $S(\rho)$ predicted by
the HF formalism of Ref.~\cite{Kho96} using the same isovector strengths $C_1$
as those used in Fig.~\ref{f1}, in comparison with the empirical points deduced
from the neutron-skin \cite{Fur02} and HI fragmentation \cite{She04} data.}
\label{f2}
\end{figure}

The \he6pn cross section given by the COSMA density \cite{Kor97} was found to
have a shape very close to that of the measured angular distribution (see
Fig.~\ref{f1}). Since the strength $N_{\rm I}/N_{\rm R}$ was fixed at 0.65, the
CC description of the \he6pn data could be improved only by fine tuning the
strength $C_1$ of the isovector part of the density dependence (\ref{g2}) of the
CDM3Y6 interaction. One can see that the best fit is achieved when $C_1$ is
about 10\% stronger than the isoscalar strength $C_0$. We have further performed
the HF calculation of asymmetric NM with the same isospin- and density dependent
CDM3Y6 interaction using the method described in Ref.~\cite{Kho96}. The density
dependence of the NM symmetry energy $S(\rho)$ obtained with the same isovector
strengths $C_1$ as those used in Fig.~\ref{f1} is shown in Fig.~\ref{f2}, and
one can deduce easily $E_{\rm sym}\approx 31\pm 1$ MeV from our HF results. This
result should be complementary to the nuclear structure studies which relate the
slope of the EOS of asymmetric NM and the associated $E_{\rm sym}$ value to the
neutron skin, a method first suggested by Alex Brown \cite{Bro00}. If one
adopts, e.g., a neutron-skin $\Delta R\approx 0.1-0.2$ fm for $^{208}$Pb then a
systematics based on the mean-field calculations (see Fig.~7 of
Ref.~\cite{Fur02}) gives $E_{\rm sym}\approx 27-31$ MeV (this value is compared
with our HF result in Fig.~\ref{f2}). The main methods to determine the neutron
skin are either the analyses of elastic ($p,p$) scattering on stable $N\neq Z$
targets \cite{Kar02,Cla03} or studies of asymmetric NM and structure of finite
nuclei \cite{Die03,Pie04}. However, the uncertainty remains still rather high
and $\Delta R$ for $^{208}$Pb nucleus was found to be ranging from 0.083-0.11 fm
\cite{Cla03} to $0.13\pm 0.03$ fm \cite{Die03} or around 0.17 fm \cite{Kar02},
and up to about 0.22 fm \cite{Pie04}. A more accurate determination of $\Delta
R$ is expected from the measurement of parity-violating electron scattering
\cite{Hor01e} and it might be used for a more precise determination of $E_{\rm
sym}$. Our result is also complementary to the nuclear reaction studies based on
the transport-model simulations \cite{Bao02,She04}. For example, a recent
antisymmetrized molecular dynamics (AMD) analysis of the HI fragmentation data
\cite{She04} obtained $S(\rho\approx 0.08$ fm$^{-3})\approx 18-20$ MeV (see
Fig.~\ref{f2}) which gives a useful information on the low-density part of
$S(\rho)$. Since the charge exchange \he6pn cross section is peaked at the most
forward angles (see Fig.~\ref{f1}), our CC analysis has probed mainly the
surface part of the FF which, in turn, is determined by the low-density part of
$v_1(E,\rho,s)$. The fact that our HF calculation reproduces quite well the
empirical half-density point of $S(\rho)$ \cite{She04} shows the reliability of
the isospin dependence of the CDM3Y6 interaction. Note that the Gogny-AS
effective $NN$ interaction used in the AMD calculation of Ref.~\cite{She04} also
gives $E_{\rm sym}\approx 31$ MeV at $\rho_0$ (see Fig.~1 of Ref.~\cite{Ono03}),
in a very close agreement with our HF result. Thus, the HF result shown in
Fig.~\ref{f2} should provide a realistic description of the EOS of asymmetric NM
for densities up to about 2 $\rho_0$.

\begin{figure}[htb]
 \vspace*{-1cm}
 \mbox{\epsfig{file=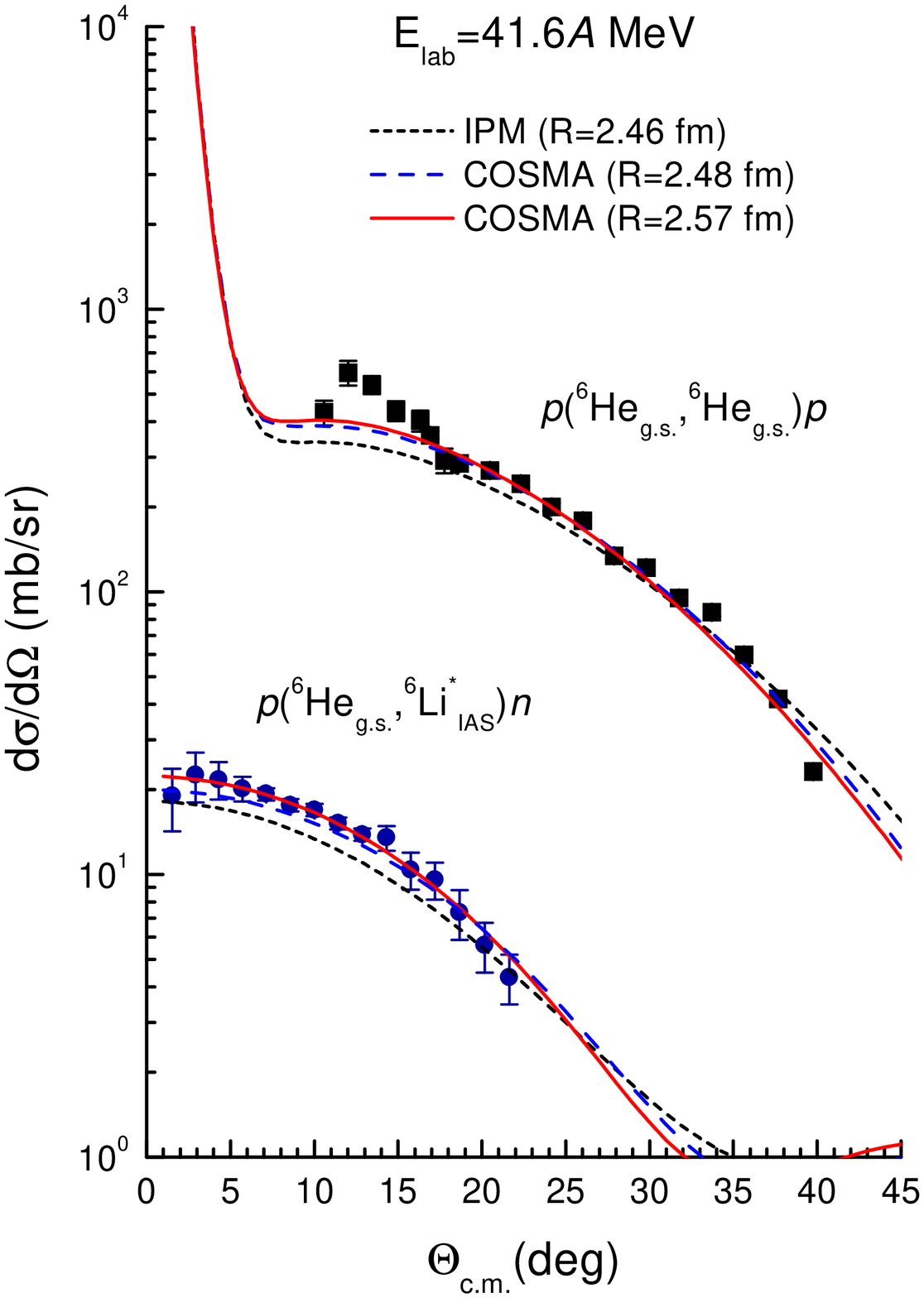,height=12cm}}\vspace*{-1cm}
\caption{CC results for the elastic scattering \pphe6 and charge exchange
\he6pn cross sections at $E_{\rm lab}=41.6A$ MeV, given by the OP and FF
obtained with 3 choices of the $^6$He$_{\rm g.s.}$ density (see text), in
comparison with the data measured by Cortina-Gil {\sl et al.}
\cite{Gil98,Gil97}.} \label{f3}
\end{figure}

\begin{figure}[htb]
 \vspace*{-1cm}
 \mbox{\epsfig{file=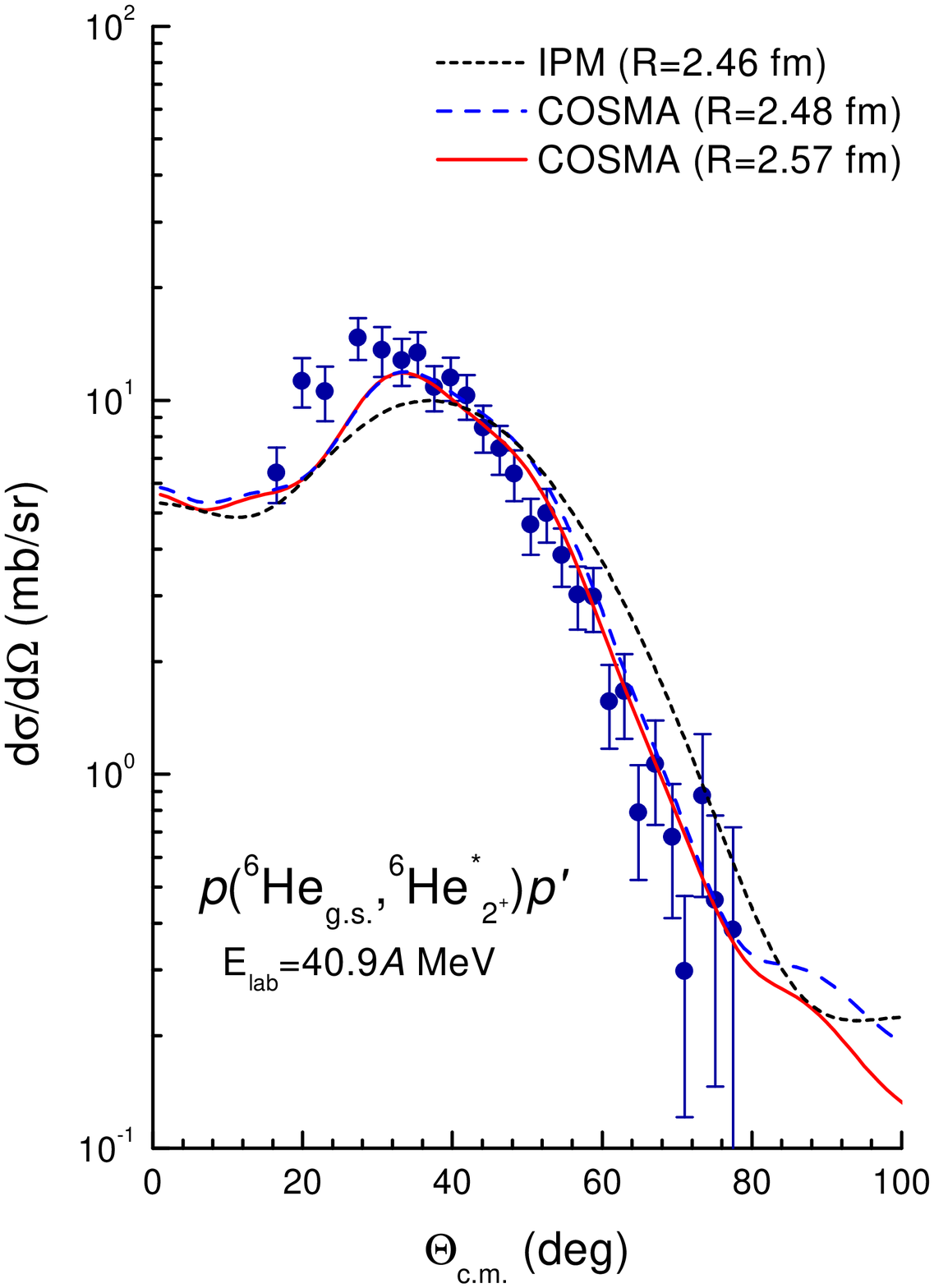,height=12cm}}\vspace*{-1cm}
\caption{DWBA results for the inelastic \p2he6 scattering to the 1.87 MeV 2$^+$
state of $^6$He at $E_{\rm lab}=40.9A$ MeV, given by the OP and FF obtained
with 3 choices of the $^6$He$_{\rm g.s.}$ density (see text), in comparison
with the data measured by Lagoyannis {\sl et al.} \cite{Lag01}.} \label{f4}
\end{figure}

Besides the conclusion on the symmetry energy, we have found further that the
measured \he6pn data are also sensitive to the halo tail of $^6$He nucleus.
Since the IAS of $^6$Li is just an isobaric analog partner of $^6$He$_{\rm
g.s.}$, the isospin symmetry implies \cite{Ara95} that this IAS of $^6$Li should
have about the same halo structure as that of $^6$He$_{\rm g.s.}$, and the
central OP's in the entrance and exit channels should be well described by
Eqs.~(\ref{g3})-(\ref{g4}). To study this effect, we have used in the folding
calculation two versions of the COSMA density \cite{Kor97} which have RMS radii
of 2.57 and 2.48 fm as well as the IPM density \cite{Kho04} which has the RMS
radius of 2.46 fm. Although the IPM density was shown \cite{Kho04} to give a
good description of the interaction cross section measured with $^6$He beam at
high energies, it was calculated in the independent particle model which does
not account for the di-neutron correlation in $^6$He. In this sense, the COSMA
densities are more accurate and should give a better description of the \he6pn
data if the IAS of $^6$Li has the same halo structure as $^6$He$_{\rm g.s.}$.
The CC results obtained with 3 different choices of the $^6$He$_{\rm g.s.}$
density are shown in Fig.~\ref{f3} and one can see that the COSMA density is
indeed more appropriate than IPM density and gives consistently good description
of both the elastic scattering and charge exchange data. With the same $N_{\rm
R(I)}$ factors used throughout these calculations, the COSMA densities with RMS
= 2.57 and 2.48 fm give $\sigma_{\rm R}=408$ and 399 mb, respectively, quite
close to the empirical value of about 400 mb. The IPM density gives $\sigma_{\rm
R}=382$ mb which is slightly smaller than 400 mb. The description of the elastic
scattering and charge exchange data by the IPM density can only be improved by
further reducing $N_{\rm I}$ factor but then $\sigma_{\rm R}$ becomes
significantly smaller than the empirical value. One can see in Fig.~\ref{f3}
that from two versions of the COSMA density the \he6pn data seem to favor that
with RMS=2.57 fm. Such a RMS radius agrees with that predicted earlier by the
three-body calculation of $^6$He which takes into account the dynamic
correlation between the $\alpha$-core and di-neutron \cite{Al96}.

In addition to the elastic \pphe6 scattering and charge exchange \he6pn
reaction, inelastic \p2he6 scattering to the 1.87 MeV 2$^+$ state of $^6$He has
also been measured \cite{Lag01} at a nearby energy of $40.9A$ MeV. This
(unbound) excitation of $^6$He has been a subject of extensive analyses of the
elastic and inelastic \phe6 scattering either in the distorted wave Born
approximation (DWBA) or CC formalism. In particular, the halo effect has been
shown to be significant in the 2$^+$ inelastic scattering channel (see Fig.~3 of
Ref.~\cite{Lag01}). Therefore, a folding analysis of the 2$^+$ inelastic
scattering data using the same \phe6 OP's as those used in Fig.~\ref{f3} would
be quite complementary to the results discussed above for the charge exchange
reaction. The real 2$^+$ inelastic FF was calculated by the folding model
\cite{Kho02,Kho03} using a simple ansatz for the transition density, where the
proton and neutron parts of the ($^6$He$_{\rm g.s.}\to ^6$He$^*_{2^+}$)
transition density are given by deforming proton and neutron parts of the
$^6$He$_{\rm g.s.}$ density with the deformation lengths $\delta_{2^+}^{(p)}$
and $\delta_{2^+}^{(n)}$ determined recently in a CC analysis of the
$^4$He($^6$He,$^6$He)$^4$He reaction \cite{KhoVoe04}. The folded 2$^+$ inelastic
FF was further scaled by the same complex factor $(1+iN_{\rm I}/N_{\rm R})$ as
that used to obtain the charge exchange FF in Eq.~(\ref{g5}). The DWBA results
obtained with three choices of the $^6$He$_{\rm g.s.}$ density are plotted in
Fig.~\ref{f4} and one can see that the best description of the \p2he6 data is
again given by the COSMA density. The deficiency of the IPM density is about the
same as that shown above in the calculated \he6pn cross sections.

Finally, we note that a folding analysis of the present charge exchange \he6pn
data has been done earlier \cite{Gil98,Vis01} using the famous JLM (complex)
G-matrix interaction. Although the real, imaginary and isovector strengths of
the JLM interaction were adjusted to the best fit of the elastic scattering and
charge exchange data, these analyses seem to be unable to give a good
description of the last data points of the measured \he6pn cross section no
matter what density distribution of $^6$He is used in the folding calculation
(see, e.g., Fig.~2 of Ref.~\cite{Vis01}). Our CC results represent, therefore,
an accurate alternative description which also provides important input for the
description of the EOS of asymmetric NM. The future measurements of the charge
exchange reactions induced by the neutron-rich beams would be very valuable in
studying the isospin dependence of the \AA interaction and making a more
reliable conclusion on the symmetry energy.

The authors thank Nicolas Alamanos and Valerie Lapoux for making the measured
cross sections available in the tabulated form. The research was supported, in
part, by Natural Science Council of Vietnam and Vietnam Atomic Energy
Commission.

\end{document}